\documentclass[useAMS,usenatbib,onecolumn]{mn2e}
\usepackage{epsfig,amssymb,amsmath,graphicx}

\newcommand{\be}{\begin{equation}}
\newcommand{\ee}{\end{equation}}

\newcommand{\rstar}{R_*}
\newcommand{\mstar}{M_*}

\title[Gravitational radiation from a magnetically deformed non-barotropic neutron star]{Gravitational wave emission from a magnetically deformed non-barotropic neutron star}
\author[A. Mastrano, A. Melatos, A. Reisenegger, and T. Akg\"{u}n]{A. Mastrano$^{1}$\thanks{E-mail:
a.mastrano@physics.unimelb.edu.au}, A.
Melatos$^{1}$\thanks{E-mail: amelatos@physics.unimelb.edu.au}, A. Reisenegger$^{2}$\thanks{E-mail: areisene@astro.puc.cl}, and T. Akg\"{u}n$^{2,3}$\thanks{E-mail: akgun@astro.cornell.edu}\\
$^{1}$School of Physics, University of Melbourne, Parkville VIC
3010, Australia\\
$^{2}$Dept. de Astronom\'{i}a y Astrof\'{i}sica, Pontificia Universidad Cat\'{o}lica de Chile, Vicu\~{n}a Mackenna 4860, Macul, Santiago, Chile\\
$^{3}$Barcelona Supercomputing Center - Centro Nacional de Supercomputaci\'{o}n, C/ Gran Capit\`{a} 2-4, Barcelona, 08034, Spain}

\begin{document}

\date{Accepted ?. Received ?; in original form ?}

\pagerange{\pageref{firstpage}--\pageref{lastpage}} \pubyear{?}

\maketitle

\label{firstpage}

\begin{abstract}

\noindent{A strong candidate for a source of gravitational waves is a highly magnetised, rapidly rotating neutron star (magnetar) deformed by internal magnetic stresses. We calculate the mass quadrupole moment by perturbing a zeroth-order hydrostatic equilibrium by an axisymmetric magnetic field with a \emph{linked poloidal-toroidal structure}. In this work, we do \emph{not} require the model star to obey a barotropic equation of state (as a realistic neutron star is not barotropic), allowing us to explore the hydromagnetic equilibria with fewer constraints. We derive the relation between the ratio of poloidal-to-total field energy $\Lambda$ and ellipticity $\epsilon$ and briefly compare our results to those obtained using the barotropic assumption. Then, we present some examples of how our results can be applied to astrophysical contexts. First, we show how our formulae, in conjunction with current gravitational wave (non-)detections of the Crab pulsar and the Cassiopeia A central compact object (Cas A CCO), can be used to constrain the strength of the internal toroidal fields of those objects. We find that, for the Crab pulsar (whose canonical equatorial dipole field strength, inferred from spin down, is $4\times 10^8$ T) to emit detectable gravitational radiation, the neutron star must have a strong toroidal field component, with maximum internal toroidal field strength $B_{\mathrm{tm}}=7\times 10^{12}$ T; for gravitational waves to be detected from the Cas A CCO at 300 Hz, $B_{\mathrm{tm}}\sim 10^{13}$ T, whereas detection at 100 Hz would require $B_{\mathrm{tm}}\sim 10^{14}$ T. Using our results, we also show how the gravitational wave signal emitted by a magnetar immediately after its birth (assuming it is born rapidly rotating, with $\Lambda\lesssim 0.2$) makes such a newborn magnetar a stronger candidate for gravitational wave detection than, for example, an SGR giant flare.}


\end{abstract}

\begin{keywords}
MHD -- stars: magnetic field -- stars: interiors -- stars: neutron -- gravitational waves
\end{keywords}

\section{Introduction}

A strong magnetic field can deform a star \citep{cf53,f54,g72,k89,pm04,hetal08}.
The ellipticity is roughly proportional to the magnetic energy \citep{hetal08,detal09}.
Therefore, it is expected that magnetars, with their intense magnetic
fields\footnote{The magnetic field strengths of magnetars are inferred from their periods
and spin-down rates, assuming dipole spin-down, but see, for example, \citet{ketal98} and
\citet{tb98} for an alternate method, applied to SGR 1806-20.}, possess significant ellipticities.
This makes them good candidates for gravitational wave sources \citep{bg96,mp05,setal05,hetal08,detal09}.

In this paper, we explore the relation between magnetic field configuration (particularly
the relative strengths of the toroidal and poloidal components) and ellipticity.
We do this by starting with a model star in hydrostatic equilibrium and impose an axisymmetric
magnetic field on it, matched to an external dipolar field to ensure a well-posed problem.
We treat the magnetic force as a perturbation on this equilibrium
and calculate the resulting small changes in pressure and density. A similar calculation of the ellipticity has been done
for purely toroidal fields in type II superconducting
neutron stars by \citet{aw08}. Theoretical arguments, as well
as the properties of pulsar glitches, strongly suggest that
superfluids (likely including superconducting protons) exist in
the interior of neutron stars (Link, Epstein, \& Baym 1993;
Yakovlev, Levenfish, \& Shibanov 1999). While superfluidity may
have a significant impact on the allowable magnetohydrodynamic
equilibria, we ignore this effect in the present study [as also
done, e.g., by Haskell et al. (2008), Lander \& Jones (2009), and
Ciolfi, Ferrari, \& Gualtieri (2010)]. Also, while our field (described in Sec. 2)
superficially resembles the `twisted torus' of Ciolfi, Ferrari, \& Gualtieri
(2010), our calculation is purely non-relativistic and, unlike
Colaiuda et al. (2008), we do not include quadrupolar (and
higher-order multipoles) fields in our calculation.

The main novelty of this paper, compared to previous works on the
subject [such as Colaiuda et al. (2008), Haskell et al. (2008),
Lander \& Jones (2009), and Ciolfi et al. (2010)], is that we do
\emph{not} require the star to be barotropic, i.e., we do
\emph{not} require it to have a one-to-one relation between
pressure and density, $p=p(\rho)$. The reason for this is that
neutron star matter is a multi-species fluid, in which different
kinds of particles co-exist, at the very least neutrons, protons,
and electrons. The relative abundances of these particles can be
adjusted by weak interactions (such as beta decays; \emph{Urca
processes} in astrophysical jargon) and diffusive processes.
However, the time-scales for these processes are much longer than
the time for the fluid to reach a hydromagnetic equilibrium (force
balance) state, which is likely to happen in a few Alfv\'en times
(of order seconds). In the process of settling to this
hydromagnetic equilibrium, each fluid element is free to move
following the relevant forces, but it conserves its composition
(relative abundances of species), which will thus not be exactly a
function of pressure and density. Once the equilibrium is reached,
the star will be radially stratified \citep{p92,rg92} to zeroth
order, but the spherical symmetry (and with it the stratification)
will be slightly perturbed by the presence of the magnetic field.

On time-scales intermediate between the Alfv\'en time and the weak
interaction and diffusion time-scales, the star will thus be in a
hydromagnetic equilibrium state in which the composition is not
strictly determined by the density or pressure, and the latter
thus do not have a one-to-one correspondence with each other. The
present study applies to this regime, considering a hydromagnetic
equilibrium state described as a spherically symmetric,
non-magnetic background in which both pressure and density depend
only on the radial coordinate, plus small, non-axisymmetric
pressure and density perturbations, caused by the presence
required to balance the Lorentz force, and which are not directly
related to each other, due to the position-dependent composition.

In this state, the particles will not be in equilibrium with
respect to weak interactions or diffusive processes. The latter
will act on longer time-scales, changing the composition and thus
the density and pressure perturbations, and thus causing a
long-term evolution of the magnetic field \citep{gr92,r09}. These
processes have been studied in some detail in a simplified,
one-dimensional model \citep{hrv08,hrv10}, which is being extended
to a more realistic, axially symmetric geometry \citep{betal11}.
When these non-ideal MHD processes become relevant, the ideal,
non-barotropic MHD equilibrium studied in this paper most likely
breaks down. At least during an important part of this later evolution, the neutron star matter is likely to be describable as two independent fluids, as done by \citet{gal11} and \citet{lag11}.


Analytically, imposing barotropy means requiring that the magnetic
force per unit mass is expressible as a gradient
\citep{p56,m65,aw08}. Relaxing this constraint {\bf for the
reasons explained above} enables us to explore a greater range of
equilibria \citep{r09}.\footnote{There are other mechanisms which
can deform a neutron star, such as `mountains' on the elastic
crust, formed by wavy electron-capture layers \citep{ucb00,hja06},
rotation, or crustal shear stresses \citep{ps74,c02}. In this
paper, we ignore these other mechanisms, so as to concentrate on
the magnetic deformation.}

In Section 2, we describe our model in detail, starting with the
zeroth-order hydrostatic equilibrium and perturbing it
magnetically. In Section 3, we derive a simple formula relating
the ellipticity $\epsilon$ to the magnetic energy, the
poloidal-to-total magnetic energy ratio, and the mass and radius
of the star, for steady states with a parabolic density profile or
a polytropic equation of state with index $n=1$. We also compare
our results to those obtained by \citet{hetal08}, who assumed
barotropy. The formula for $\epsilon$ has several astrophysical
applications. In Section 4, we discuss these and compare our work
to previous studies in the literature, such as \citet{c02}. We
apply our results to gravitational wave emission from a newly
born, rapidly rotating magnetar. We discuss the possibility of
detecting the gravitational-wave signal from a newly born magnetar
in the Virgo cluster and relate its detectability (quantified by
the signal-to-noise ratio) to the magnetic field configuration,
emphasizing the poloidal-toroidal energy ratio. We then apply our
formulae, together with the current upper limits from the Laser
Interferometer Gravitational-Wave Observatory (LIGO), to constrain
the strength of the toroidal fields of the Crab pulsar and the
Cassiopeia A central compact object. We summarize our results and
discuss further applications and refinements of our idealised
model in Section 5.

\section{Hydromagnetic perturbation}

Let $(r,\theta,\phi)$ be the spherical polar coordinates, with $r$ expressed in units of the stellar radius $\rstar$ (so that it is dimensionless). As in \citet{aetal11}, we choose a magnetic field configuration such that: (1) $\nabla\cdot{\bf{B}}=0$ is satisfied; (2) the magnetic field is axially symmetric around the $z$-axis; (3) the poloidal component of the field is continuous with an external dipole field, which vanishes as $r\rightarrow\infty$; (4) there are no surface currents (the tangential, toroidal component vanishes at the surface); (5) the toroidal component is confined to the region of closed poloidal field lines around the neutral line (on which the poloidal component vanishes); and (6) the current density remains finite and continuous everywhere in the star. We write the magnetic field inside the star in the form pioneered by \citet{c56}, as a sum of toroidal and poloidal components,

\be {\bf{B}}=B_0[\eta_p\nabla\alpha(r,\theta)\times\nabla\phi + \eta_t\beta(\alpha)\nabla\phi],\ee
where $\eta_p$ and $\eta_t$ are dimensionless parameters which define the relative strengths of the poloidal and toroidal components, respectively. The flux function $\alpha(r,\theta)$ is taken to be $f(r)\sin^2\theta$. Note that this particular form of $\alpha(r,\theta)$ is only applicable when we try to match our field to an external dipole field; other multipoles will have different $\theta$-dependences. The radial dependence of $\alpha$ is defined by

\be f(r)=\frac{35}{8}\left(r^2-\frac{6r^4}{5}+\frac{3r^6}{7}\right).\ee
The function $f(r)$ is postulated to be of this form to ensure that the field described by Eqs. (1)--(3) is continuous with a dipole field outside the star, that there are no surface currents, and that the current density is finite at the origin [for a more thorough derivation, see \citet{aetal11}]. The function $\beta(\alpha)$ is chosen as

\be
\beta(\alpha) =
   \begin{cases}
      (\alpha-1)^2& ,{\textrm{ }}\alpha\geqslant 1,\\
      &\\
      0& ,{\textrm{ }}\alpha< 1,
   \end{cases}
\ee
which ensures that the toroidal field is confined to the region where $\alpha$ exceeds $1$, the value taken by $\alpha$ at $r=1$, $\theta=\pi/2$, with the current density going continuously to zero at this boundary.

\begin{figure}
\centerline{\epsfxsize=10cm\epsfbox{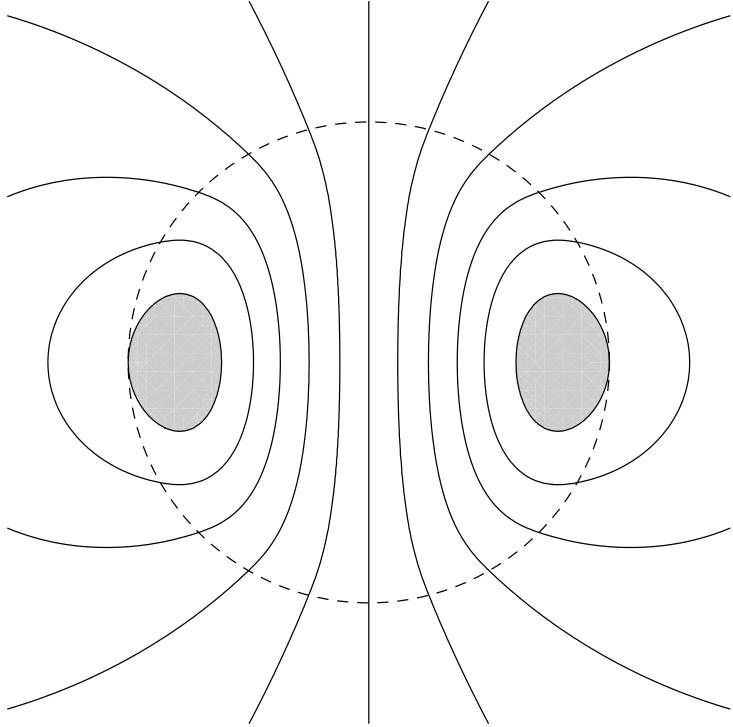}}
 \caption{A diagram of the magnetic field lines described by Eqs. (4)--(6). The surface of the star is represented by the dashed circle. The toroidal magnetic component is confined to the shaded region and describes a circle around the $z$-axis. The poloidal field lines vanish at the neutral line. All field lines lie on surfaces of constant $\alpha$.}
 \label{field}
\end{figure}

The magnetic structure described by Eqs. (1)--(3) is drawn schematically in Fig. \ref{field}. The field lines lie on surfaces of constant $\alpha$. The toroidal component is completely restricted to the shaded region, which is bounded by the surface $\alpha(r,\theta)=1$. Note that both the toroidal and poloidal components are continuous across $\alpha=1$, so that there are no surface currents at $\alpha=1$, either inside the star or on the surface at the equator. The poloidal component inside the $\alpha\geqslant1$ region (the shaded region) lies on nested tori around the neutral line, with the strength of the poloidal component continuously decreasing towards the neutral line. Outside the star, the poloidal component is matched to a dipole field, consistent with $\alpha(r,\theta)\propto \sin^2\theta$.

If the magnetic field is treated as a perturbation on the background\footnote{Care must be taken, however, that neither the poloidal nor the toroidal component of the field approaches the virial limit,

\begin{equation*} B_{\mathrm{virial}} \sim 1.4\times 10^{14} \left(\frac{\mstar}{1.4 M_\odot}\right)\left(\frac{\rstar}{10^4 \mathrm{m}}\right)^{-2} \textrm{T},\end{equation*}
where the magnetic field energy is equal to the gravitational binding energy \citep{lp07}. The exact value of this limit depends on actual field structure. In fact, \citet{r09} showed that the upper limit is set by stable stratification at $\sim 10^{13}$ T.}, we can write the force balance equation to first order in $B^2/\mu_0 p$ as\footnote{Throughout this paper, we use SI units. 1 T $= 10^{4}$ G.}

\be (1/\mu_0)(\nabla\times{\bf{B}})\times{\bf{B}}=\nabla\delta p+\delta\rho\nabla\Phi,\ee
where the Cowling approximation has been taken (i.e., $\delta\Phi=0$). Note that we do \emph{not} require the density perturbation $\delta\rho$ to be a function of the pressure perturbation $\delta p$ (the barotropic assumption). Therefore, the magnetic field is not required to satisfy the Grad-Shafranov equation.


Substituting the field given by Eqs. (1)--(3) into Eq. (4) gives

\be -\frac{B_0^2}{\mu_0 r^2\sin^2\theta}(\eta_p^2\nabla\alpha\hat{\Delta}\alpha+\eta^2_t\beta\nabla\beta)=\nabla\delta p+\delta\rho\nabla\Phi,\ee
with the operator

\be\hat{\Delta}=\frac{\partial^2}{\partial r^2}+\frac{\sin\theta}{r^2}\frac{\partial}{\partial\theta}\left(\frac{1}{\sin\theta}\frac{\partial}{\partial\theta}\right).\ee
The $\theta$-component of the right-hand side of the force balance equation (5) gives $\partial(\delta p)/\partial\theta$, which is integrated with respect to $\theta$ to give $\delta p$. Then $\delta\rho$ is obtained from the radial component of (5). Thus, we write
\be \delta {p}=\frac{-B_0^2\alpha}{\mu_0 r^2\sin^2\theta}\left[\eta_p^2\hat{\Delta}\alpha+2\eta_t^2\left(\frac{\alpha^3}{3}-\frac{3\alpha^2}{2}+3\alpha-\ln\alpha+C_0\right)\right],\ee
\be \delta {\rho}\frac{d\Phi}{dr}=\frac{B_0^2}{\mu_0}\left[\eta_p^2\alpha\frac{\partial}{\partial r}\bigg(\frac{\hat{\Delta}\alpha}{r^2\sin^2\theta}\bigg)+2\eta^2_t\left(\frac{\alpha^3}{3}-\frac{3\alpha^2}{2}+3\alpha-\ln\alpha+C_0\right)\frac{\partial}{\partial r}\left(\frac{\alpha}{r^2\sin^2\theta}\right)\right],\ee
where $C_0$ is a constant, needed to ensure $\delta p$ is continuous across $\alpha=1$. In this particular case, where the field is defined by Eqs. (1)--(3), we find $C_0=-11/6$. Equations (7)--(8) are applicable to the region where $\alpha\geqslant 1$; to obtain similar expressions for the $\alpha < 1$ region (where the toroidal field vanishes), one simply sets $\eta_t=0$.

An arbitrary function of radius $h(r)$ can be added to $\delta p$. This is equivalent to changing the background profile by a spherically symmetric function, hence it does not affect the mass quadrupole moment, which is the focus of this paper. For computational simplicity, this function is set to zero (so that $\delta p$ vanishes along the $z$-axis). Note that, while both the Lagrangian and Eulerian pressure perturbations vanish at the surface, the Eulerian density perturbation [Eq. (8)] does not (only the Lagrangian perturbations need to be zero at the perturbed surface).

We integrate the squares of the toroidal and poloidal components of the magnetic field over all space (including the vacuum, to $r\rightarrow\infty$) to obtain their energies. We can write the total magnetic energy $E_{\mathrm{m}}$ as


\be E_{\mathrm{m}}=(a_p\eta_p^2+a_t\eta_t^2)\frac{\pi B_0^2 \rstar^3}{\mu_0},\ee
with the dimensionless constants $a_p = 8.48$ and $a_t=1.65\times 10^{-5}$. We can then express the ratio of the energy of the poloidal magnetic field component to the total magnetic energy as $\Lambda$

\be \Lambda = \frac{\int_V dV {\bf{B}}_{\mathrm{p}}^2/\mu_0}{\int_V dV{\bf{B}}^2/\mu_0}=\frac{\eta^2_p}{\eta^2_p+q\eta^2_t},\ee
with $q=1.95\times 10^{-6}$. By definition, one has $0\leqslant\Lambda\leqslant 1$, where $\Lambda=0$ and $\Lambda=1$ represent a star with a purely toroidal and a purely poloidal field, respectively.

In Fig. \ref{lambda}, we plot $\delta p$ and $\delta \rho (d\Phi/dr)$ for various values of $\Lambda$, in units of $B_0^2/\mu_0$, taking $\eta_p=1$ without loss of generality. In the first column, we plot $\mu_0\delta p /B_0^2$ (solid curves) and $\mu_0\delta\rho (d\Phi/dr) /B_0^2$ (dashed curves) along $\theta=\pi/2$ as a function of $r$ for some values of $\Lambda$, to show how the internal toroidal magnetic field gradually changes the pressure and density distributions. In the purely poloidal case, $\delta p$ has a maximum at $r=0.54$. As the toroidal field strength increases (i.e., as $\Lambda<1$), local minima for $\delta p$ and $\delta \rho (d\Phi/dr)$ develop straight away at $r=0.78$ (where the toroidal part of the magnetic field has a maximum), which eventually become global minima. In a sense, the pressure and density perturbations which are solely due to the poloidal field (the peak at $r=0.54$) are gradually overwhelmed by the perturbations caused by the strengthening toroidal field component (the peak at $r=0.78$). In the second and third columns, we plot contours of $\mu_0\delta p /B_0^2$ and $\mu_0\delta\rho (d\Phi/dr) /B_0^2$ in a meridional quadrant, to show their dependence on the polar angle. Note how the contours are clustered around the new minimum at $r=0.78$, created by the increasingly stronger toroidal component, as $\Lambda$ decreases.

\begin{center}
\begin{figure*}
\begin{tabular}{cc}
\begin{tabular}{c}
\mbox{$\Lambda=1$}
\end{tabular}

&
\begin{tabular}{c}
\mbox{}

\end{tabular}

\\
\begin{tabular}{c}
\includegraphics[height=42mm]{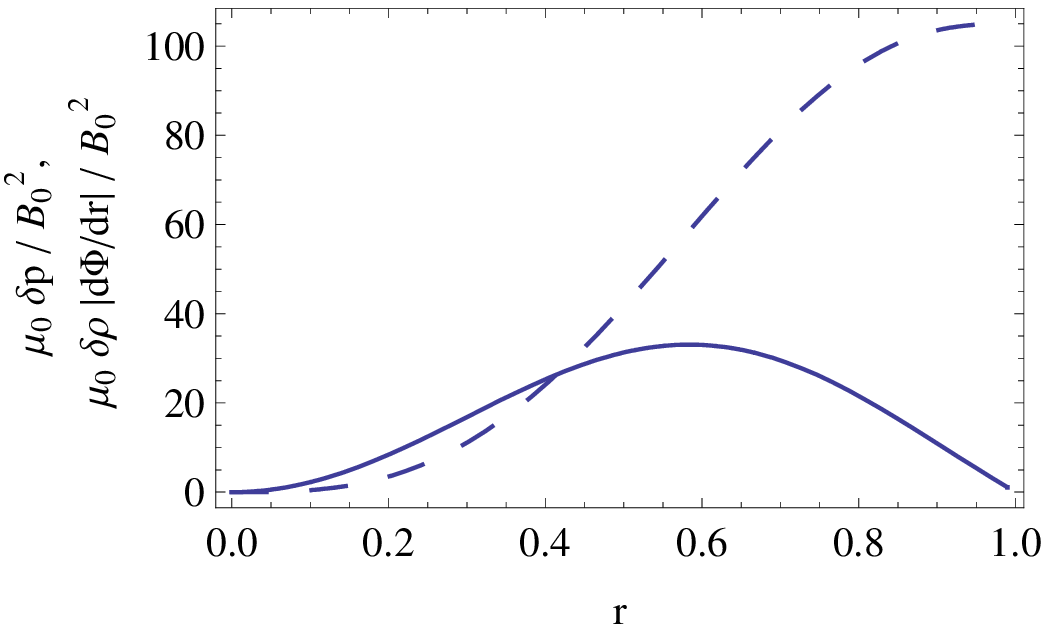}
\end{tabular}

&
\begin{tabular}{c}
\includegraphics[height=45mm]{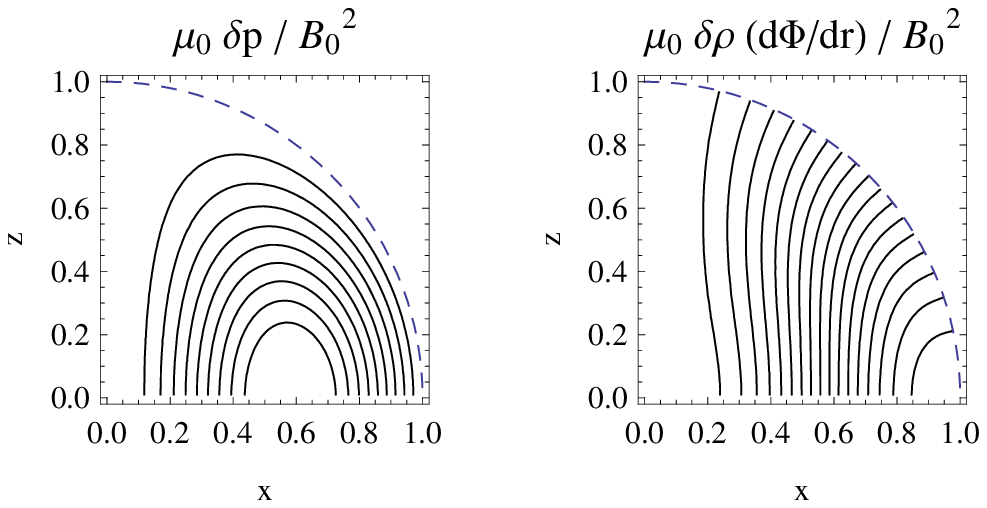}
\end{tabular}
\\

\begin{tabular}{c}
\mbox{$\Lambda=0.9$}
\end{tabular}

&
\begin{tabular}{c}
\mbox{}
\end{tabular}
\\

\begin{tabular}{c}
\includegraphics[height=42mm]{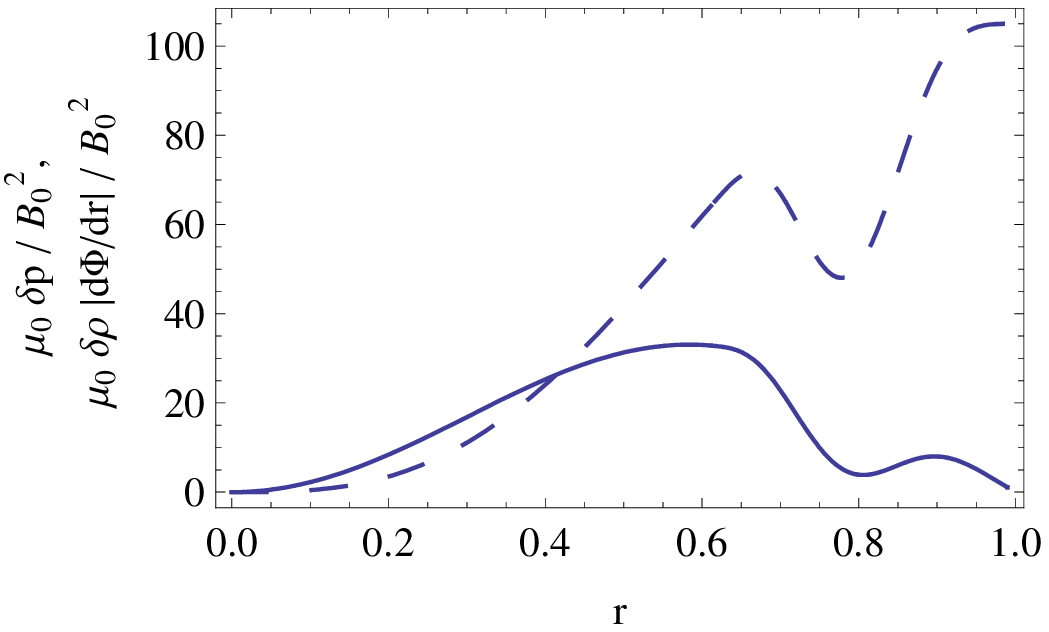}
\end{tabular}

&
\begin{tabular}{c}
\includegraphics[height=45mm]{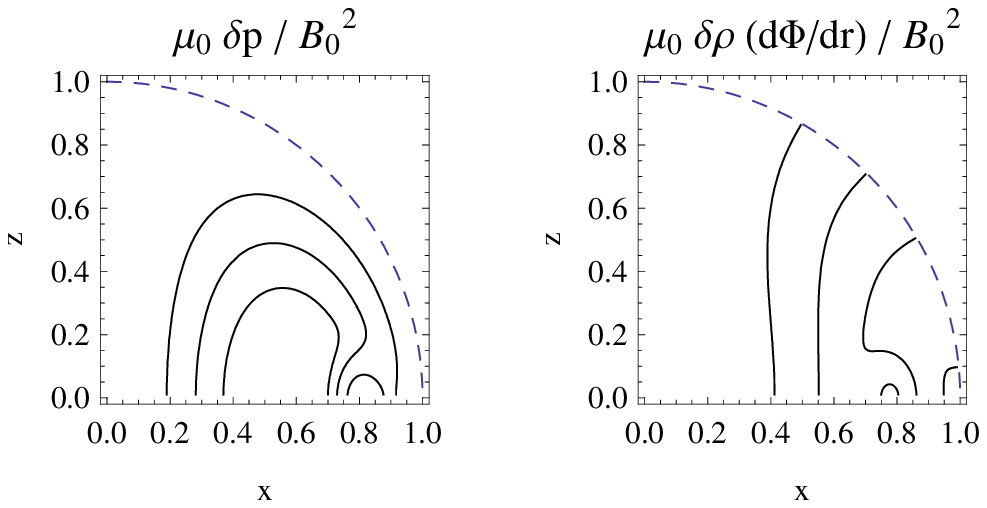}
\end{tabular}
\\

\begin{tabular}{c}
\mbox{$\Lambda=0.8$}
\end{tabular}

&
\begin{tabular}{c}
\mbox{}
\end{tabular}
\\

\begin{tabular}{c}
\includegraphics[height=42mm]{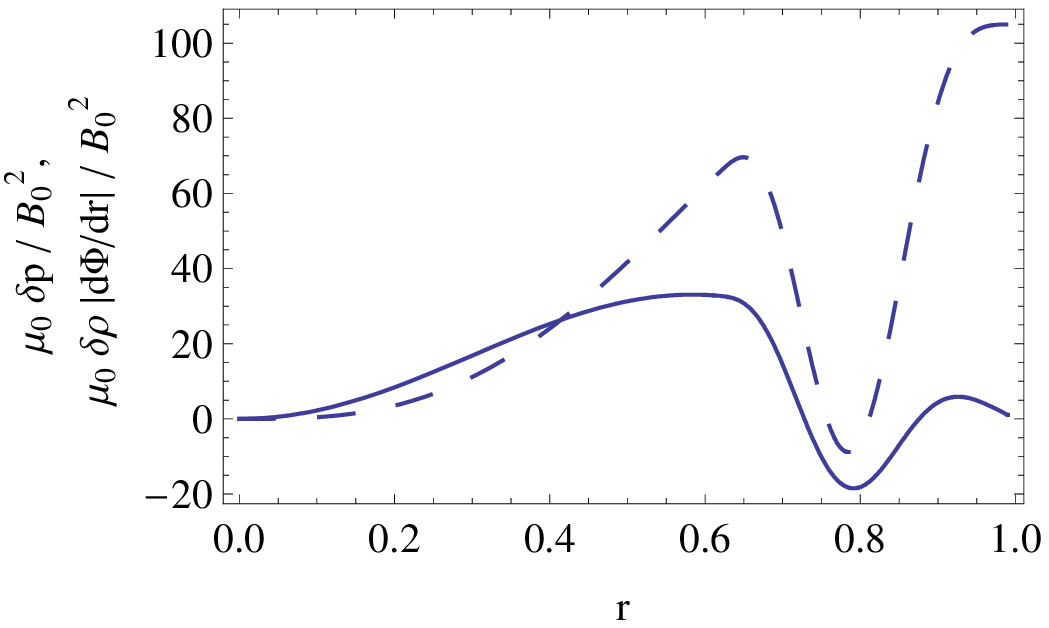}
\end{tabular}

&
\begin{tabular}{c}
\includegraphics[height=45mm]{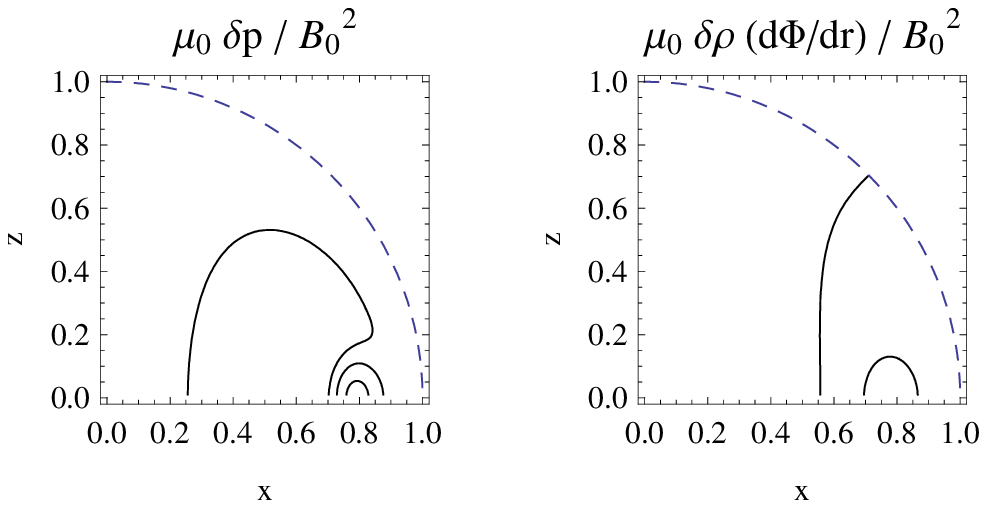}
\end{tabular}
\\


%
%
%


\begin{tabular}{c}
\mbox{$\Lambda=0.1$}
\end{tabular}

&
\begin{tabular}{c}
\mbox{}
\end{tabular}
\\

\begin{tabular}{c}
\includegraphics[height=42mm]{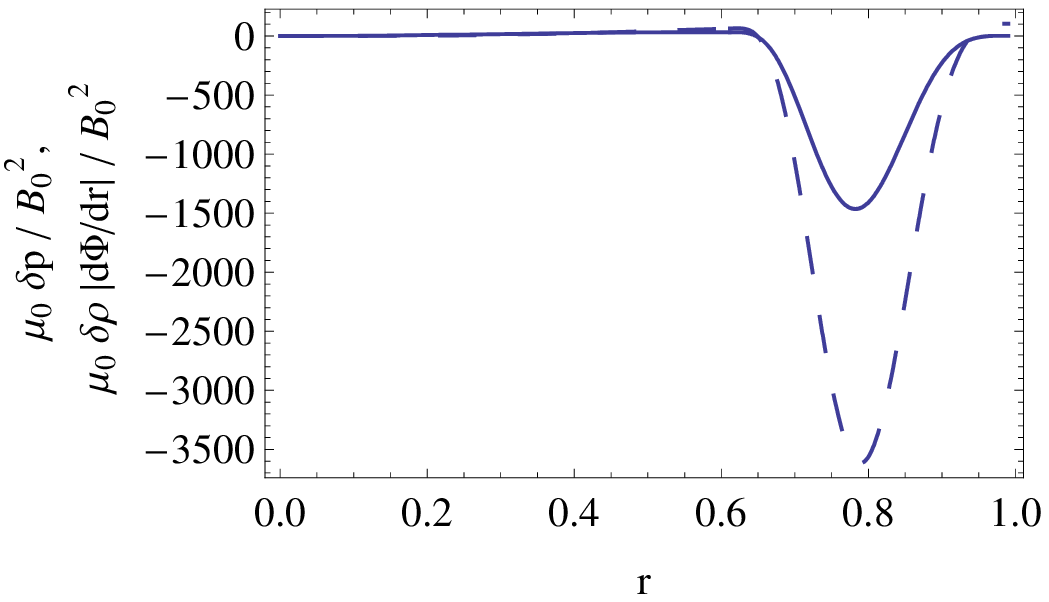}
\end{tabular}

&
\begin{tabular}{c}
\includegraphics[height=45mm]{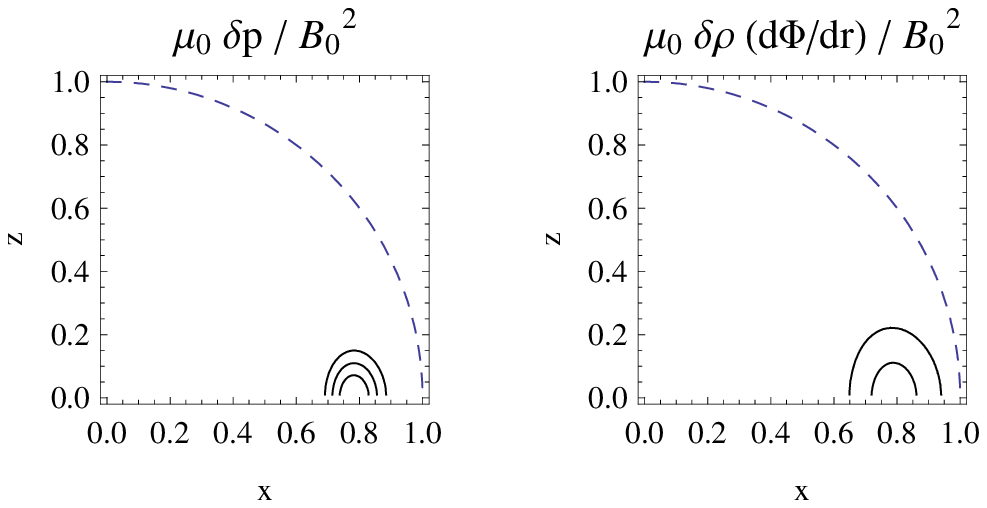}
\end{tabular}

\end{tabular}

\caption{(Left column) Pressure and density perturbations $\mu_0\delta p /B_0^2$ (solid curve) and $\mu_0\delta\rho |d\Phi/dr| /B_0^2$ (dashed curve) versus normalised radius $r$, at $\theta=\pi/2$, with $\eta_p=1$ and various values of $\Lambda$. (Middle and right columns) Contour plots of $\mu_0\delta p /B_0^2$ and $\mu_0\delta\rho |d\Phi/dr| /B_0^2$, respectively, on the meridional quadrant. Note that contour levels differ between panels, they are auto-scaled to emphasize the structure of $\mu_0\delta p /B_0^2$ and $\mu_0\delta\rho |d\Phi/dr| /B_0^2$. The dashed curves represent the surface of the star.}
\end{figure*}
\label{lambda}
\end{center}

\section{Magnetically induced deformation}

\subsection{Ellipticity}

The gravitational wave strain generated by a rotating rigid body is directly proportional to its ellipticity. Hence, our first step is to calculate the ellipticity of the star due to the magnetic configuration given by Eqs. (1)--(3). The ellipticity $\epsilon$ is defined as

\be \epsilon = \frac{I_{zz}-I_{xx}}{I_0},\ee
where $I_0$ is the moment of inertia of the spherical star. The moment-of-inertia tensor is given by

\be I_{jk}=\rstar^5\phantom{+}\int_V dV [\rho(r) + \delta\rho(r,\theta)](r^2\delta_{jk}-x_jx_k),\ee
which allows us to rewrite (11) as

\be \epsilon = \pi\rstar^5 I_0^{-1}\int_V dr d\theta\delta\rho(r,\theta) r^4\sin\theta(1-3\cos^2\theta).\ee

\begin{center}
\begin{figure*}
\begin{tabular}{cc}
\begin{tabular}{c}
\mbox{(a)}
\end{tabular}

&
\begin{tabular}{c}
\mbox{(b)}
\end{tabular}

\\
\begin{tabular}{c}
\includegraphics[scale=0.7]{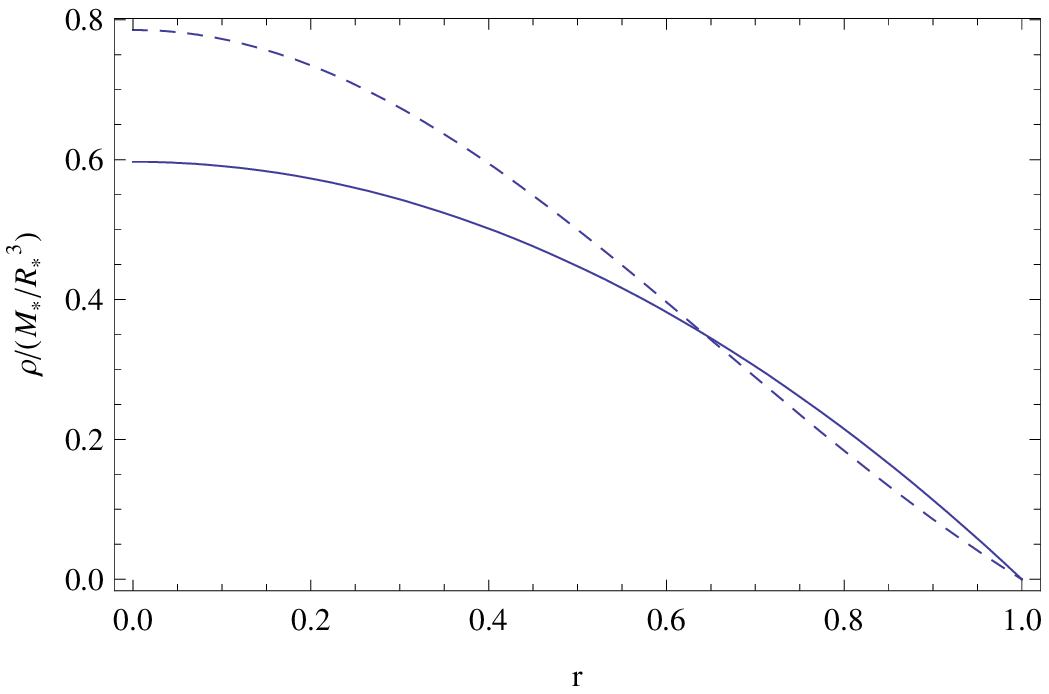}
\end{tabular}

&
\begin{tabular}{c}
\includegraphics[scale=0.7]{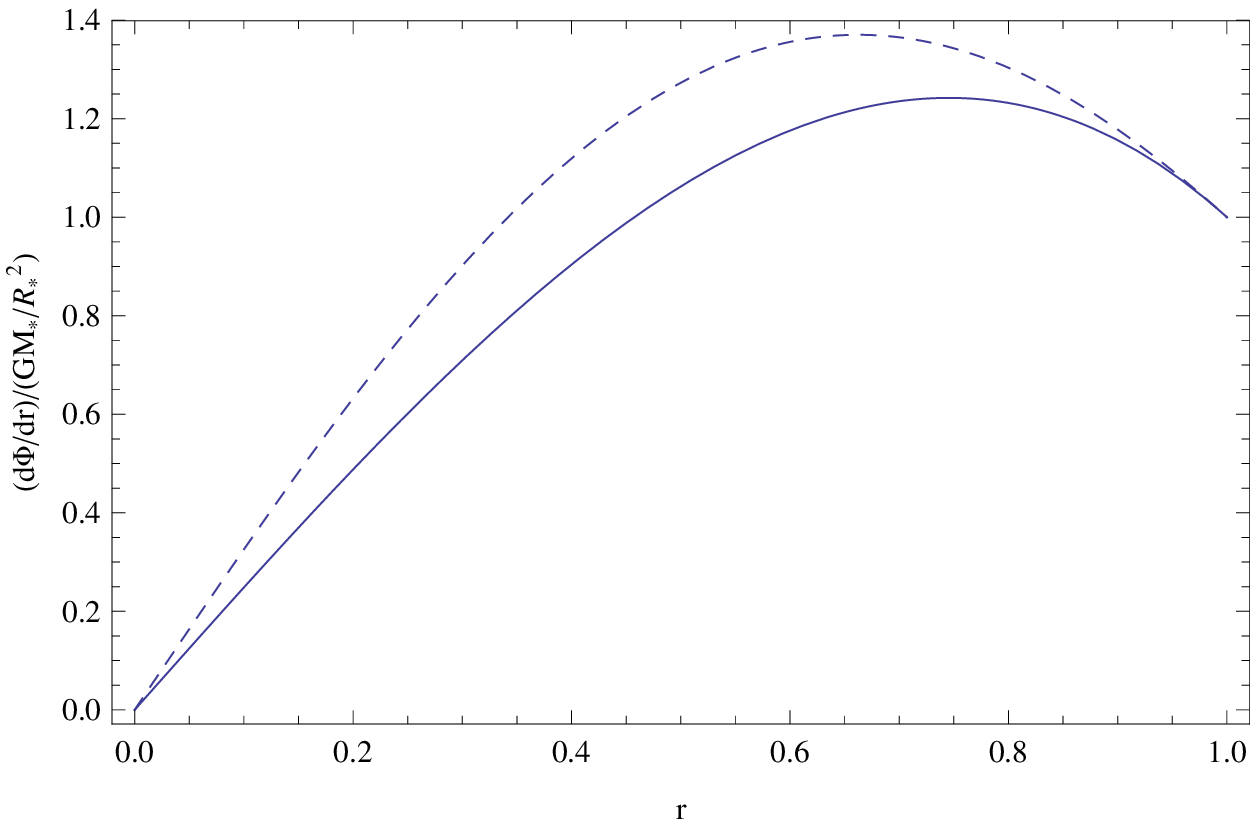}
\end{tabular}

\end{tabular}

\caption{(a) Density profiles for the unmagnetized parabolic steady state given by Eq. (14) (solid curve) and the unmagnetized $n=1$ polytropic steady state given by Eq. (18) (dashed curve) in units of $\mstar/\rstar^3$; (b) gravitational acceleration $d\Phi/dr$ as a function of $r$ for the unmagnetized parabolic steady state given by Eq. (15) (solid curve) and the unmagnetized $n=1$ polytropic steady state given by Eq. (20) (dashed curve) in units of $G\mstar/\rstar^2$.}
\label{rhophicomp}
\end{figure*}
\end{center}

Roughly speaking, the ellipticity induced by the magnetic field is proportional to the magnetic energy, which determines $\delta\overline{\rho}$, as is evident from Eq. (8). We wish to know how the details of the magnetic configuration, specifically $\Lambda$, control $\epsilon$. Below, we apply our model to two different unmagnetized steady states. We show the density and $d\Phi/dr$ profiles for these unmagnetized steady states in Fig. \ref{rhophicomp} (a) and (b), respectively.

\subsection{Parabolic steady state}

In the absence of a magnetic field, the star is in a spherically symmetric hydrostatic equilibrium. Here, we consider the simple density profile \citep{aetal11}


\be \rho =\rho_c (1-r^2),\ee
where $\rho_c=15\mstar/(8\pi\rstar^3)$ is the density at the centre. The gravitational acceleration (i.e., the gradient of the gravitational potential $\nabla\Phi$) corresponding to this steady state is

\be \frac{d\Phi}{dr} = \frac{G\mstar}{2\rstar^2} r(5-3r^2),\ee
where $\mstar$ and $\rstar$ denote the mass and radius of the star, respectively and $G$ is Newton's gravitational constant. From the gravitational acceleration, we obtain the pressure profile

\be p=p_c \left(1-\frac{5r^2}{2}+2r^4-\frac{r^6}{2}\right),\ee
where $p_c=15G\mstar^2/(16\pi\rstar^4)$ is the pressure at the centre. We emphasize that this is a particular, simple choice of density profile, chosen to render the following calculations tractable; it is not motivated directly by physical arguments or observations.

For this steady state, Eq. (13) gives

\be \epsilon=5.98\times10^{-6}\left(\frac{B_{\mathrm{surface}}}{5\times 10^{10}\textrm{T}}\right)^2\left(\frac{\mstar}{1.4M_\odot}\right)^{-2}\left(\frac{\rstar}{10^4\textrm{m}}\right)^4\left(1-\frac{0.389}{\Lambda}\right),\ee
where $B_{\mathrm{surface}}$ is the surface magnetic field strength at the equator; we have $B_{\mathrm{surface}}=\eta_p B_0$ in our formulation.

\subsection{Polytropic steady state}

For comparison, we also consider the $n=1$ polytropic star, whose zeroth-order density and pressure configurations are given by \citep{aw08,detal09}

\be \rho=\frac{\rho_c \sin(\pi r)}{\pi r}, \phantom{++++}\rho_c=\frac{\pi\mstar}{4\rstar^3},\ee
\be p=k\rho^2, \phantom{++++}k=\frac{2G\rstar^2}{\pi}.\ee
The gradient of the gravitational potential is

\be \frac{d\Phi}{dr} = \frac{G\mstar}{\pi\rstar}\left(\frac{\sin\pi r}{r^2}-\frac{\pi\cos\pi r}{r}\right).\ee

For this steady state, Eq. (13) gives

\be \epsilon=6.262\times10^{-6}\left(\frac{B_{\mathrm{surface}}}{5\times 10^{10}\textrm{T}}\right)^2\left(\frac{\mstar}{1.4M_\odot}\right)^{-2}\left(\frac{\rstar}{10^4\textrm{m}}\right)^4\left(1-\frac{0.385}{\Lambda}\right).\ee

\begin{figure}
 \includegraphics{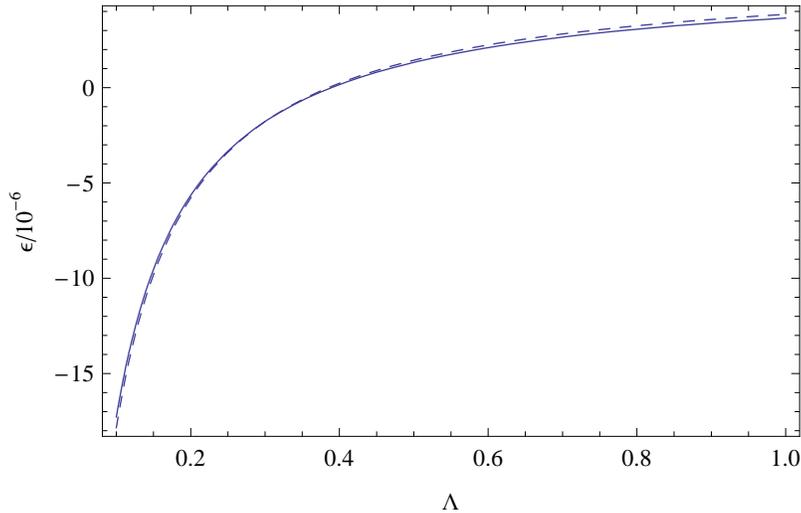}
 \caption{Ellipticity $\epsilon$ (in units of $10^{-6}$) versus $\Lambda$, the ratio of poloidal magnetic field energy to total magnetic field energy, for $\eta_p=1$, $B_0=5\times 10^{10}$ T, $\rstar =10^4$ m, $\mstar=1.4 M_\odot$. Note that $\epsilon$ is proportional to $B_0^2\rstar^4\mstar^{-2}$. The solid and dashed curves are for a parabolic [Eq. (15)] and $n=1$ polytrope [Eq. (19)] steady-state density profile, respectively.}
 \label{elambdauspolyt}
\end{figure}

We plot $\epsilon$ as a function of $\Lambda$ for an $n=1$ polytropic star as the dashed curve in Fig. \ref{elambdauspolyt}. Note that, like in the parabolic steady state, the ellipticity in the $n=1$ polytropic steady state case also vanishes at $\Lambda_{\mathrm{c}}\approx 0.39$. Equations (17) and (21) show that the $n=1$ polytrope and the parabolic steady states give ellipticities within 5\% of each other. For consistency and analytic simplicity, throughout the rest of this paper, we use the parabolic steady state to calculate ellipticity.

\subsection{Comparison with \citet{hetal08}}

We now briefly compare our result to that obtained by \citet{hetal08}. The latter authors postulate an $n=1$ polytropic star which also obeys the barotropic condition. This restricts the form of the magnetic field, as the (first-order) density and pressure perturbations are required to be proportional to each other. The magnetic field in \citet{hetal08} is characterized by the discrete eigenvalue $\lambda$ (not to be confused with our $\Lambda$), as described by Eqs. (B.6)--(B.10) of that paper. Furthermore, \citet{hetal08} do not make the Cowling approximation. Otherwise, the two methods are fundamentally similar: begin with an unmagnetized steady state in hydrostatic equilibrium, apply an axisymmetric magnetic field, assume the departure from the hydrostatic equilibrium caused by the magnetic field is small enough to be treatable as a perturbation, then calculate the perturbations to pressure and density due to the applied magnetic field.

\citet{hetal08} present their results for $\epsilon$ as a function of $\lambda$ (and, hence, volume-averaged absolute values of the toroidal and poloidal field strengths $\overline{B}_t$ and $\overline{B}_p$) in their Table 1 [see also the associated Erratum \citep{hetal09}]. We calculate $\epsilon$ for our field configuration using Eq. (21) and compare our $\epsilon$ to those obtained by \citet{hetal08} for the same values of internal poloidal-to-total field energy ratios in Table 1 (where we also show the corresponding values of $\Lambda$). We emphasize here that a direct comparison between the results of \citet{hetal08} and ours is not entirely appropriate, because we use very different field configurations. However, by using Table 1, we simply wish to illustrate how the resulting ellipticities are very different, even when using the same poloidal-to-total field energy ratios.



\begin{center}
\begin{figure*}
\begin{tabular}{c}
\begin{tabular}{c}
\mbox{(a)}
\end{tabular}
\\

\begin{tabular}{c}
\includegraphics[scale=1.1]{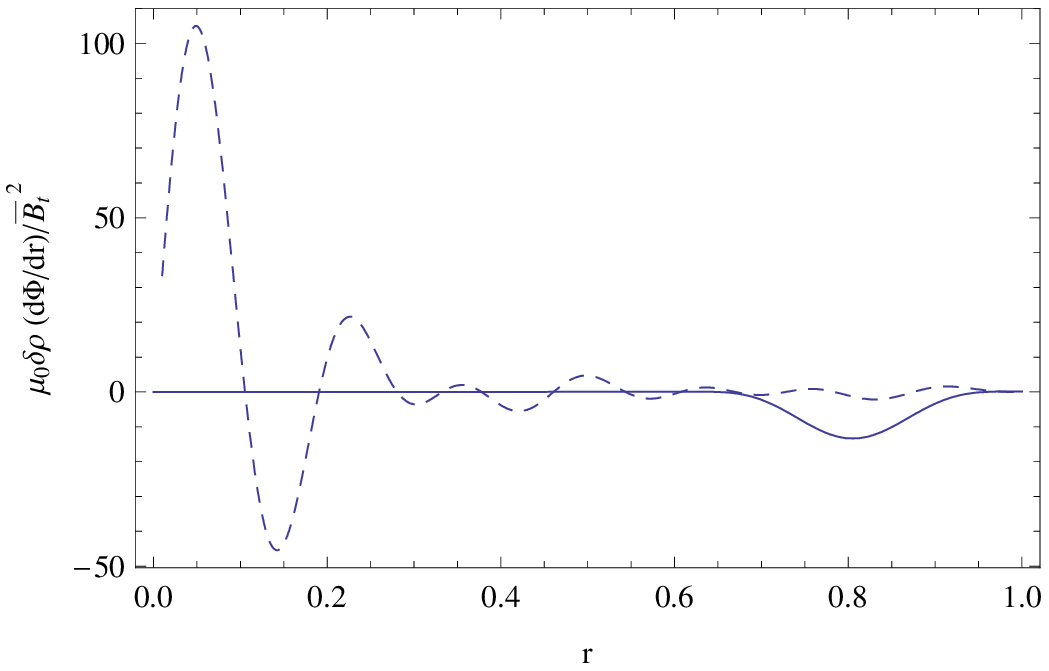}
\end{tabular}

\\

\begin{tabular}{c}
\mbox{(b)}
\end{tabular}
\\

\begin{tabular}{c}
\includegraphics[scale=1.2]{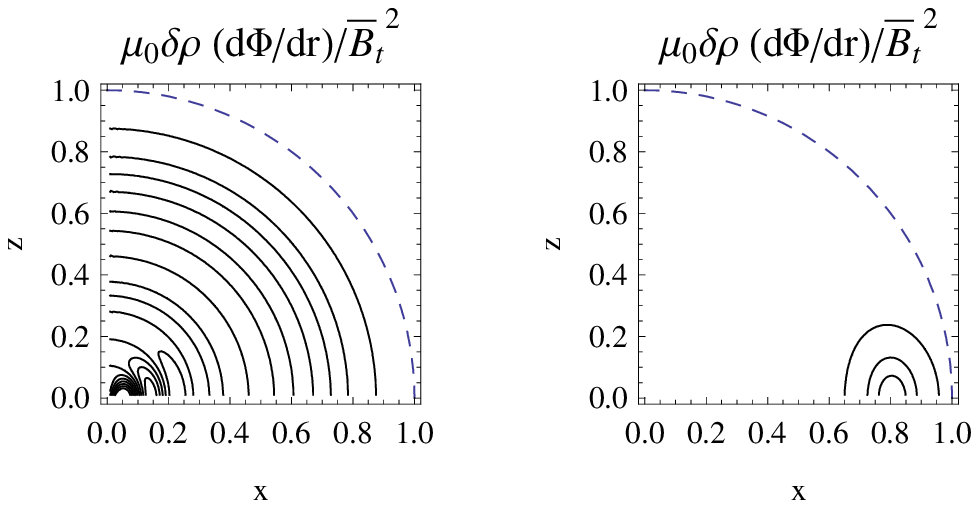}
\end{tabular}

\\

\begin{tabular}{c}
\mbox{(c)}
\end{tabular}
\\

\begin{tabular}{c}
\includegraphics[scale=1.05]{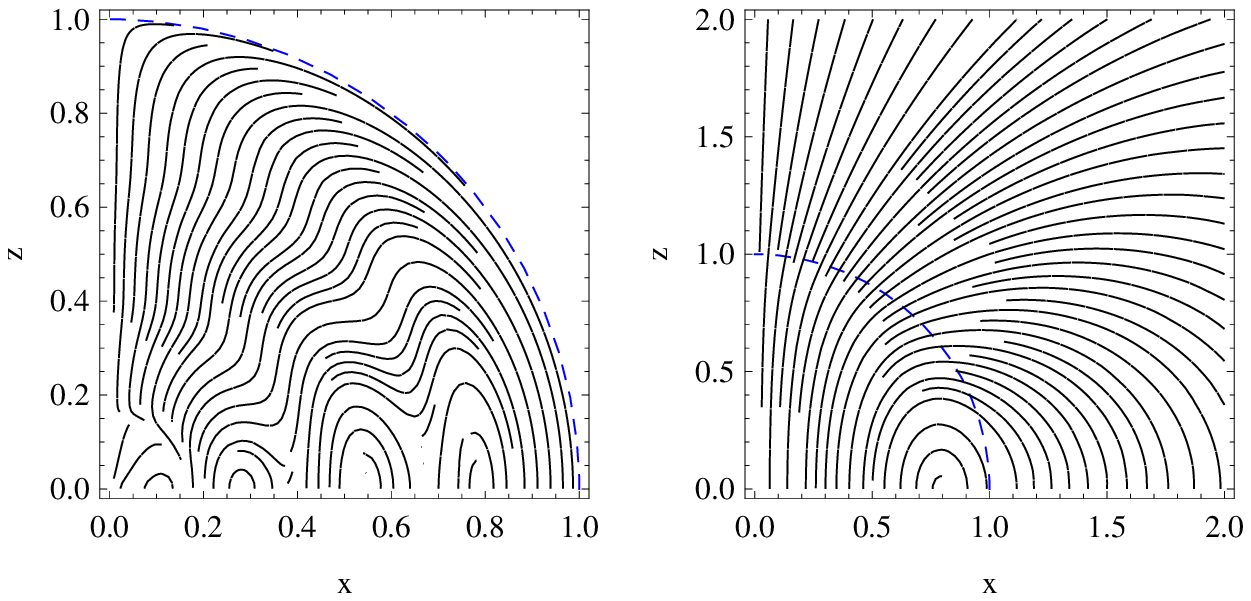}
\end{tabular}

\end{tabular}

\caption{Comparison of the field configuration of \citet{hetal08} and that of this paper, taking $\Lambda=1.77\times 10^{-2}$ [corresponding to $\lambda=7.459$ in the notation of \citet{hetal08}]: (a) $\mu_0 \delta\rho(d\Phi/dr)/\overline{B}_t^2$ as a function of radius $r$ (normalised to the stellar radius), at the equator for the fields described by Eqs. (4.1)--(4.3) (solid curve) and by Eqs. (B6)--(B10) of \citet{hetal08}; (b) contour plots of $\mu_0 \delta\rho(d\Phi/dr)/\overline{B}_t^2$ for the field used by \citet{hetal08} (left panel) and our field (right panel); (c) internal poloidal field lines for the field used by \citet{hetal08} (left panel) and our field (right panel).}
\label{haskcomp}
\end{figure*}
\end{center}

As \citet{hetal08} did, we find that the star is always prolate for the chosen field configuration and toroidal field strengths, and becomes more prolate as the toroidal field component grows. We generally find larger values of $\epsilon$ for smaller $\lambda$ and smaller $\epsilon$ for larger $\lambda$. We caution the reader again that we and \citet{hetal08} use different magnetic field configurations:

\begin{itemize}
\item our toroidal field is concentrated within an equatorial torus, while that of \citet{hetal08} is not confined to this region, varies smoothly across the stellar interior (with the number of oscillations controlled by $\lambda$), and vanishes at the origin, the stellar surface, and the $z$-axis;
\item our poloidal field is continuous with an external dipole field, whereas that of \citet{hetal08} vanishes at the surface;
\item the comparisons we make in this section are between states with the same internal poloidal-to-total field energy ratios.
\end{itemize}
While our methods also differ in that we take the Cowling approximation whereas \citet{hetal08} do not, we contend that it is the fundamental difference between field configurations that causes the discrepancies in $\epsilon$ seen in Table 1. To illustrate this further, we show the contour plot of $\mu_0\delta \rho (d\Phi/dr)$ [i.e., the radial component of the force balance equation Eq. (4)] for $\lambda=7.459$ as Fig. \ref{haskcomp}. Note how the radial component of the internal poloidal field lines of the configuration used by \citet{hetal08} vanishes at the surface of the star (Fig. \ref{haskcomp}.c, left panel), unlike our field configuration, where the radial component is continuous with a dipole field outside the star (Fig. \ref{haskcomp}.c, right panel). {As an aside, in the context of accreting neutron stars, Haskell et al. (2006) have shown that abandoning the Cowling approximation alters the possible maximum surface deformation (due to accreted mountains), and ellipticity, by a factor of 0.5--3 (depending on the density profile chosen). While this is a different physical phenomenon than the one discussed here (i.e., deformation due to accretion rather than due to internal magnetic fields), this result indicates that taking the Cowling approximation may change our ellipticities by a factor of order unity. Of course, a full calculation needs to be undertaken to determine if this is truly the case, and, if it is indeed so, whether the Cowling approximation increases or lowers ellipticities.}


\begin{table*}
 \centering
 \begin{minipage}{170mm}
  \caption{Comparison between the ellipticities $\epsilon$ obtained by \citet{hetal08} ($\epsilon_\mathrm{H}$, fourth column) and by using Eq. (21) ($\epsilon_{21}$, fifth column), as a function of the eigenvalue $\lambda$ and volume-averaged toroidal field strength $\overline{B}_t$, with the volume-averaged poloidal field strength $\overline{B}_p$ kept constant at $10^8$ T. We show the corresponding poloidal-to-total field energy ratio $\Lambda$ in the third column to make contact with the rest of the calculations and results in this paper.}
  \begin{tabular}{@{}lccccc@{}}
  \hline
     $\lambda$ &$\overline{B}_t$ (T)& $\Lambda$&$\epsilon_\mathrm{H}$&$\epsilon_{21}$&$\epsilon_\mathrm{H}/\epsilon_{22}$\\
\hline 2.362 & $3\times 10^8$&0.116 & $-9.6\times 10^{-13}$& $-5.4\times 10^{-12}$& 0.18\\
3.408 & $4.2\times 10^8$&$6.29\times 10^{-2}$ &$-4.2\times 10^{-12}$ & $-1.2\times 10^{-11}$&0.35\\
7.459 &$8.1\times 10^8$ & $1.77\times 10^{-2}$&$-6.6\times 10^{-11}$ & $-4.8\times 10^{-11}$&1.38\\
25.488 & $2.7\times 10^9$& $1.62\times 10^{-3}$&$-3.5\times 10^{-10}$ & $-5.5\times 10^{-10}$&0.64\\
33.491 & $3.5\times 10^9$& $9.65\times 10^{-4}$&$-6.6\times 10^{-10}$ & $-9.3\times 10^{-10}$&0.71\\
58.495 & $9.3\times 10^9$&$1.37\times 10^{-4}$&$-5.6\times 10^{-8}$&$-6.95\times 10^{-9}$&8.06\\
318.499 & $4.8\times 10^{10}$&$5.14\times 10^{-6}$&$-2\times 10^{-6}$&$-1.85\times 10^{-7}$&10.81\\

\hline
\end{tabular}
\end{minipage}
\end{table*}

\section{Gravitational-wave astrophysical limits on $\epsilon$ and $\Lambda$}


Equations (10) and (17) allow us to use observational limits on ellipticity to constrain the strength of the internal toroidal field component. There are many possible contexts in which we can apply these ideas and results. In this section, we look at two illustrative applications:

\begin{itemize}
\item isolated pulsars with or without timing solutions;
\item birth of isolated magnetars.
\end{itemize}

\subsection{Upper limit of gravitational radiation from the Crab pulsar}

Recent Laser Interferometer Gravitational-Wave Observatory (LIGO) Science Run 5 observations of the Crab pulsar have put limits on its ellipticity \citep{aetal08,aetal09}. Using Eq. (17) and these data, we can constrain the strength of the surface and internal magnetic fields of the Crab pulsar. In Table 2, we list the lower limits of $\Lambda$ based on the latest upper limits of detection from LIGO and the planned Einstein Telescope in its B \citep{hcf08} or C \citep{hetal10} configuration. From the fundamental definition of $\Lambda$ [Eq. (10)] and the magnetic field defined by Eqs. (1)--(3), we can express the maximum of the internal toroidal field $B_{\mathrm{tm}}$ as a function of $B_0$ and $\Lambda$ to facilitate comparison:

\be B_{\mathrm{tm}}=18.2 B_0 \left(\frac{1}{\Lambda}-1\right)^{1/2}.\label{btm}\ee
To give a numerical example, for $B_{\mathrm{surface}}$ (the surface dipole field at the equator, equivalent to $\eta_pB_0$ in our formulation) corresponding to the theoretical value for pure dipole braking (namely $3.8\times 10^{8}$ T) and $\Lambda=0.4$ (i.e., $B_{\mathrm{tm}}=8.4\times 10^9$ T) we find $\epsilon=8.7\times 10^{-12}$, which is much lower than the LIGO upper limit. It is clear that, if indeed the surface magnetic field strength is $B_{\mathrm{surface}}\sim 10^8$ T, as inferred from electromagnetic spin down, detection of gravitational waves from the Crab pulsar is only possible if the star is prolate with a very strong toroidal component ($\Lambda\lesssim 10^{-6}$, corresponding to $B_{\mathrm{tm}}\gtrsim 1.82\times 10^{12}$ T). In other words, the poloidal field alone cannot deform the Crab pulsar enough to be detectable as a gravitational wave source by LIGO and the planned Einstein B or C; a strong internal toroidal field is required to deform the star into a high-$|\epsilon|$ prolate shape. Indeed, if gravitational waves were to be detected from the Crab pulsar, the analysis presented in this paper offers a method to infer the strength of this internal toroidal field.

\begin{table*}
 \centering
 \begin{minipage}{170mm}
  \caption{Inferred minimum $\Lambda_{\mathrm{min}}$ of the Crab pulsar, using Eq. (17), based on the upper limits of gravitational wave strain ($h_0$) from various gravitational wave detectors \citep{aetal08,aetal09,b10}. We note that, in the case of the Crab pulsar, because of its (relatively) weak magnetic field strength, significant wave strain can only be achieved if the star is highly prolate.}
  \begin{tabular}{@{}lccc@{}}
  \hline
     Detector &$|h_0|$ (Hz$^{-1/2}$)& $|\epsilon|$ & $\Lambda_{\mathrm{min}}$\\
\hline
LIGO & $2.6\times 10^{-25}$ & $1.4\times 10^{-4}$ & $9.6\times 10^{-7}$\\
Einstein B & $3.0\times 10^{-25}$ & $1.6\times 10^{-4}$ & $8.4\times 10^{-7}$\\
Einstein C & $5.0\times 10^{-25}$ & $2.7\times 10^{-4}$ & $5.0\times 10^{-7}$\\

\hline
\end{tabular}
\end{minipage}
\end{table*}

\subsection{The case of Cassiopeia A Central Compact Object}

The supernova that gave birth to the supernova remnant Cassiopeia A (henceforth Cas A) was probably observed in 1680 \citep{a80}, although the central compact object (CCO), hypothesized to be a neutron star, was only positively identified as a bright X-ray point source in 1999 \citep{t99}, making this one of the youngest neutron stars in our Galaxy \citep{hh09}. Because the location of the CCO is well known, it makes for an attractive target for a directed gravitational wave search with LIGO \citep{wetal08}.

While the location of the CCO is well known, it has no detectable pulse in any electromagnetic band, rendering it impossible to measure its rotation frequency\footnote{However, due to the matched filtering method used by LIGO, detection of gravitational waves would automatically give rotation rate.}, frequency derivative, and magnetic field \emph{a priori} \citep{wetal08}. However, recent analysis has shown that a neutron star model with a low magnetic field can explain the Cas A observations \citep{hh09}. After fitting several different atmospheric models (namely blackbody, pure hydrogen, helium, carbon, and nitrogen) to the measured X-ray spectrum, Ho and Heinke (2009) concluded that the Cas A CCO is most likely a neutron star with a carbon atmosphere, radius 8--17 km, mass 1.5--2.4 $M_\odot$, and upper limit of $10^{7}$ T for the magnetic field. For this set of physical parameters, our model predicts $\epsilon\sim 10^{-15}$ if the internal field is purely poloidal. Therefore, we expect the star to be prolate with a large toroidal component (a very low $\Lambda$) if it is to approach the LIGO upper limits estimated by \citet{wetal08} and \citet{w10}.

\begin{figure}
 \includegraphics[scale=1.0]{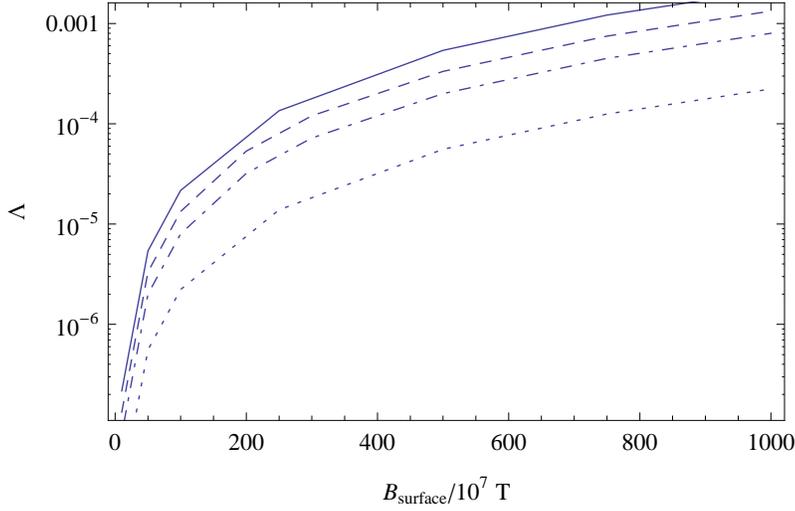}
 \caption{Contours of constant gravitational wave frequencies plotted on the $\Lambda$--$B_\mathrm{surface}$ plane, indicating the upper limits on $B_\mathrm{surface}$ and lower limits on $\Lambda$ for each given frequency (solid curve: 300 Hz; dashed curve: 200 Hz; dashed-dotted curve: 150 Hz; dotted curve: 100 Hz) for the Cas A CCO, inferred from the most recent LIGO upper limits \citep{w10} and the analysis by Ho and Heinke (2009). We take $\mstar=1.5 M_\odot$, $R=10^4$ m. For the case of $\mstar=2.4 M_\odot$, $R=1.7\times 10^4$ m, $\Lambda$ is higher by a factor of 3.26 for each given $B_p$ and gravitational wave frequency.}
 \label{casap}
\end{figure}

Based on the Ho and Heinke (2009) conclusions and the most recent upper limits on the ellipticity set by LIGO, obtained from indirect upper limits on intrinsic gravitational wave strain \citep{w10}, we can now set lower limits on $\Lambda$ of the Cas A CCO (i.e., upper limits on the toroidal field strength). Assuming $\mstar=1.5 M_\odot$, $R=10^4$ m, we take the upper limits on $\epsilon$ for four different putative gravitational wave frequencies $f$ (twice the spin frequency): 100, 150, 200, and 300 Hz. We then plot the lower limit on $\Lambda$ as a function of the (unknown) surface field strength in Fig. \ref{casap}. The curves in Fig. \ref{casap} thus indicate the $\Lambda$ needed at various $B_\mathrm{surface}$ for detection by LIGO at 100 (dotted curve), 150 (dashed-dotted curve), 200 (dashed curve), and 300 Hz (solid curve); the region to the left and above each curve generates $\epsilon$ that is below the sensitivity limits of LIGO. The LIGO upper limits for $\epsilon$ at lower $f$ are higher [e.g., $\epsilon=3.6\times 10^{-4}$ at $f=100$ Hz, see Fig. 4 of \citet{w10}], hence the smaller limits on $\Lambda$ at lower frequencies seen in Fig. \ref{casap}.

While the corresponding limits for the internal magnetic field strength may seem to be very large (even approaching $B_{\mathrm{virial}}$ at low $f$), note that, given the lack of other information to constrain $B_{\mathrm{tm}}$, these are merely \emph{upper limits}. From Fig. \ref{casap} and Eq. (22), we see that our model predicts that, for gravitational waves to be detected from Cas A at 300 Hz, the CCO must have $B_{\mathrm{tm}}\geqslant 4\times 10^{13}$ T (if it has mass 1.5 $M_\odot$ and radius $10^4$ m) or $B_{\mathrm{tm}}\geqslant 2\times 10^{13}$ T (if it has mass 2.4 $M_\odot$ and radius $1.7\times 10^4$ m); for gravitational wave detection at 100 Hz, the CCO must have $B_{\mathrm{tm}}\geqslant 1.2\times 10^{14}$ T (if it has mass 1.5 $M_\odot$ and radius $10^4$ m) or $B_{\mathrm{tm}}\geqslant 6.5\times 10^{13}$ T (if it has mass 2.4 $M_\odot$ and radius $1.7\times 10^4$ m).


\subsection{Detectability of gravitational wave signals from newly-formed magnetars in the Virgo cluster}


According to certain evolutionary scenarios, magnetars are hypothesized to be born with periods $P \sim 1$ ms \citep{td93,yb98,detal09}. If so, significant amounts of gravitational radiation may be emitted during a magnetar's early days \citep{ds07,detal09}. Gravitational radiation is maximal when the symmetry axis is orthogonal to the spin axis, and the spin down of a neutron star (including a magnetic contribution with arbitrary braking index $n$) is then given by \citep{st83,m97,wetal08,detal09,cetal10}

\be \dot{P}=\frac{2^{n}\pi^3R^6B_{\mathrm{surface}}^2}{3\mu_0Ic^3P^{n-2}}\left(\frac{\pi R}{c}\right)^{n-3}+\frac{2^9\pi^4GI\epsilon^2}{5c^5P^3}\ee
and the rate of rotational energy loss is given by

\be \dot{E}_{\mathrm{rot}}=-\frac{4\pi^2 I\dot{P}}{P^3}=-\frac{2^{n+2}\pi^5R^6B_{\mathrm{surface}}^2}{3\mu_0c^3P^{n+1}}\left(\frac{\pi R}{c}\right)^{n-3}-\frac{2^{11}\pi^6GI^2\epsilon^2}{5c^5P^6},\ee
where, again, $B_{\mathrm{surface}}$ is the surface dipole field at the equator (equivalent to $\eta_pB_0$), and $n$ is the braking constant.

\begin{figure}
 \includegraphics{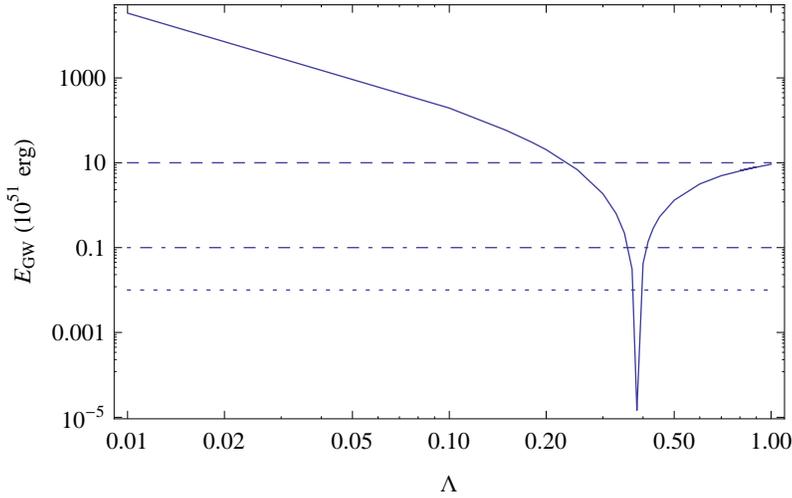}
 \caption{Gravitational wave spin-down energy extracted from a rapidly rotating, newly born magnetar (initial period 1 ms, surface dipole field $5\times 10^{10}$ T at the equator, radius 10 km, mass 1.4 $M_\odot$, distance 10 Mpc) in the first 100 s as a function of the poloidal-to-total-magnetic energy ratio $\Lambda$. Also shown for comparison are the gravitational wave energy upper limits from the SGR 1806-20 giant flare (dotted line: white noise burst waveform 100 ms 100--1000 Hz; dashed line: circularly polarised waveform 200 ms 2590 Hz) \citep{aetal08} and from the simulated SGR 1900+14 storm search (dashed-dotted line: circularly polarised waveform 200 ms 1590 Hz) \citep{ketal09}. The latter sources are astrophysically unrelated to a newly born magnetar; they are included as rough guides to the sort of signal levels detectable at present. Note that we show the vertical axis in units of $10^{51}$ ergs to facilitate comparison with \citet{tetal04}.}
 \label{egwpb}
\end{figure}

It has been suggested by \citet{tetal04} that a newly born magnetar can be a good candidate for gravitational wave detection during the first 100 s of its life. Firstly, using our model, assuming that the magnetar is born with initial period of 1 ms, surface equatorial dipole field of $5\times 10^{10}$ T, radius of 10 km, mass of 1.4 $M_\odot$, $\Lambda =0.8$,\footnote{This value is the upper limit of $\Lambda$ for a stable star with axisymmetric magnetic field \citep{b09}.} we find that the effects of gravitational spin down on the period are negligible compared to the electromagnetic spin down. Spin down purely due to gravitational wave emission can only increase the newborn magnetar's period by $10^{-3}$ \%, whereas pure electromagnetic spin down can more than double the period within these first 100 s [cf. top left panel of Fig. 5 of \citet{tetal04}]. Even though the effect on spin down is marginal, the power emitted in gravitational radiation during these first 100 s may still be significant. Thus, we plot the gravitational rotational energy extracted from this newly born, rapidly rotating magnetar in the first 100 s as Fig. \ref{egwpb} [cf. bottom left panel of Fig. 5 of \citet{tetal04}]. Note the close quantitative correspondence between our strongly poloidal case ($0.9\leqslant\Lambda\leqslant 1$) and their Case C. Our Fig. \ref{egwpb} also shows that energy output cannot distinguish between the $0.9\leqslant\Lambda\leqslant 1$ case and $0.22\leqslant\Lambda\leqslant 0.25$. We also plot the gravitational wave energy upper limits from SGR 1806-20 giant flare observations and SGR 1900+14 simulated storm search \citep{ketal09} for rule-of-thumb comparison (despite their astrophysically different natures). It is clear that a newly-born, fast-rotating magnetar with $\Lambda \leqslant 0.2$ can be a significant source of gravitational waves, stronger than an SGR giant flare, and thus a promising target for gravitational wave searches during its first 100 s.

\citet{c02} and \citet{setal05} [further refined by \citet{detal09}] approximated the optimal matched-filter signal-to-noise $(S/N)$ ratio of newly born magnetars in the Virgo cluster by integrating the instantaneous signal strain amplitude over the frequency range $f_i\leqslant f \leqslant f_f$ ($f_i$ is the initial, `birth' spin frequency of the magnetar and $f_f$ is the final frequency, when we assume the most significant gravitational wave emission ceases) and averaging over detector antenna pattern. For Advanced LIGO at frequencies 0.5--2 kHz, they give the following formula for $S/N$:

\be S/N = \frac{4}{5}\sqrt{\frac{\pi G I}{6c^3}}\frac{\pi}{DS_0^{1/2}}\left(\frac{K_{GW}}{K_d}\right)^{1/2}\bigg(2\ln\frac{f_i}{f_f}-\ln\frac{a+f_i^2}{a+f_f^2}\bigg)^{1/2},\ee
where $S_0\approx 2.1\times 10^{53}$ Hz$^{-1}$ is the squared noise spectral density at frequencies between 0.5 and 2 kHz \citep{detal09}, $K_d=2\pi^3\rstar^6B_{\mathrm{surface}}^2/(3\mu_0Ic^3)$, $K_{GW}=32\pi^4G\epsilon^2I/(5c^5)$, $a=K_d/\pi^2K_{GW}$, and $D$ is the distance to the source.

\begin{figure}
 \includegraphics{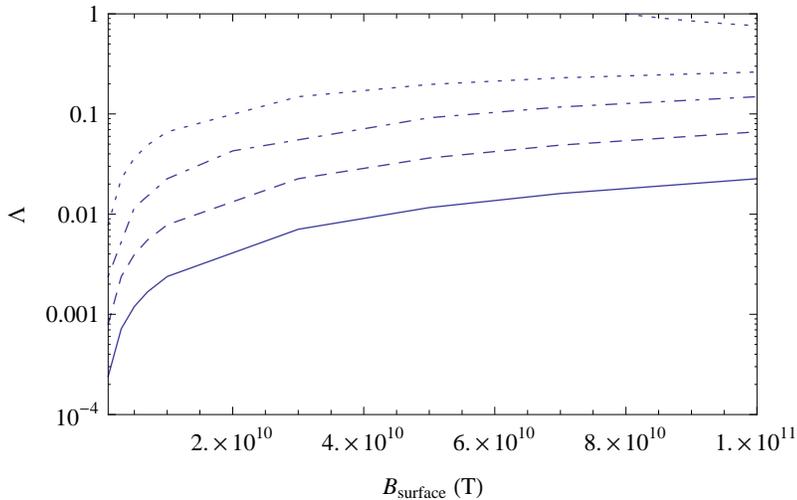}
 \caption{Relation between $\Lambda$ and $B_{\mathrm{surface}}$ for given values of signal-to-noise ratio $(S/N)$, for a $\mstar=1.4$ $M_\odot$, $\rstar=10^4$ m source located 20 Mpc away, rotating with initial spin period 0.97 ms and final spin period 10 s. Solid curve: $S/N=10$; dashed curve: $S/N=3$; dashed-dotted curve: $S/N=1$; dotted curve: $S/N=0.3$.}
 \label{spern097}
\end{figure}

Taking $D=20$ Mpc, $f_i=1.03\times 10^3$ Hz, $f_f= 0.1$ Hz, $\mstar=1.4$ $M_\odot$, $\rstar=10^4$ m as an example, also assuming the source is an orthogonal rotator, we plot $\Lambda$ as a function of $B_{\mathrm{surface}}$ [using Eq. (17) to relate $\Lambda$ and $\epsilon$] for $S/N=0.3$, 1, 3, and 10 in Fig. \ref{spern097}. Note that we obtain generally smaller $S/N$ compared to the predictions made by \citet{detal09}. This is because \citet{detal09} used the \citet{c02} formula, which assumes that the toroidal field dominates and the poloidal field is negligible, to calculate their $\epsilon$, whereas we use Eq. (17) to calculate ours, which is derived self-consistently in Sections 2 and 3 (without neglecting the poloidal component, which makes our star less prolate). According to our model, significant detectability ($S/N\approx 10$) for $B_{\mathrm{surface}}=5\times 10^{10}$ T is only possible when $\Lambda\lesssim 0.01$, when the star is prolate and the maximum toroidal field strength is $\approx 3500$ times the poloidal field strength at the surface; when the field is purely poloidal $(\Lambda=1)$, we predict $S/N\approx 0.18$ only.

\section{Conclusions}

Neutron stars in general are not expected to be barotropic \citep{p92,rg92}. They possess observable external dipole fields and internal linked poloidal-toroidal fields \citep{bn06,b09}. Non-barotropic assumption allows us freedom in constructing such equilibria.\footnote{The \emph{stability} of these equilibria is not considered in this paper; it is the subject of another paper in preparation \citep{aetal11}.} In this paper, we model a highly magnetised neutron star as a fluid sphere in hydrostatic equilibrium perturbed by an axisymmetric, poloidal-toroidal magnetic field, whose toroidal component is entirely confined within the star and whose poloidal component is continuous with a dipole field outside of the star. Our model differs from those of previous authors [e.g., \citet{hetal08}] in that we do not assume the star to be barotropic. We then self-consistently calculate the ellipticity caused by density perturbations in a magnetar and derive a simple relation between ellipticity $\epsilon$ and the ratio of poloidal to total magnetic field energy $\Lambda$. In principle, equation (17) allows one to infer the internal magnetic structure of a neutron star from observable quantities (namely, ellipticity, mass, surface field strength, and radius). For a magnetar with surface dipole field strength of $B_{\mathrm{surface}}=5\times 10^{10}$ T, radius 10 km, mass 1.4 $M_\odot$, with a purely poloidal field configuration ($\Lambda =1$), we predict an ellipticity of $3\times10^{-6}$ and expect such a star to be oblate.

Applying our results to specific situations, we show how current gravitational-wave non-detections can be used to put constraints on the strengths of the internal toroidal magnetic fields of the Crab pulsar and the Cas A CCO. Then, we show that a newly born, fast-rotating magnetar can emit gravitational radiations with energies of $\sim 10^{45}$ J (if $\Lambda =1$ or 0.2) during the first 100 s of its life (on a par with the SGR 1806-20 giant flare) and that a newborn millisecond magnetar in the Virgo cluster can be a stronger candidate for a detectable gravitational wave source than an SGR giant flare. We also predict the signal-to-noise ratio for a source located 20 Mpc away and relate this ratio to $\Lambda$.


In this paper, we ignore physical effects such as crustal stiffness, the presence of superfluids, and rotational deformations. These effects warrant further investigation. Furthermore, a fuller model is required to explain details of magnetic reconfiguration. In conjunction with works such as \citet{b09}, \citet{bn06}, and \citet{bs04,bs06}, our work can also be extended to investigate the implications of magnetic field configurations on various internal aspects of a neutron star, such as the equation of state.

\section*{Acknowledgments}

We thank the referee, Lars Samuelsson, for his thoughtful and timely comments that allowed us to improve and clarify the manuscript.

This work is supported by the Melbourne University International Postgraduate Research Scholarship, the Albert Shimmins Memorial Fund, FONDECYT Postdoctoral Grant 3085041, FONDAP Center for Astrophysics (15010003), Basal Center for Astrophysics and Associated Technologies (PFB-06/2007), Proyecto L\'{i}mite VRI 2010-15, and FONDECYT Regular Grants 1060644 and 1110213.

\bsp \label{lastpage}

\end{document}